\documentclass[12pt]{article}
\usepackage{epsf}

\begin{document}

\title{
The structure of the exact effective action and the quark confinement
in MSSM QCD.}

\author{{Stepanyantz K.V.}
\thanks{E-mail: stepan@theor.phys.msu.su}}

\maketitle

\begin{center}
{\em Moscow State University, Moscow, Russia}
\end{center}

\begin{abstract}
An expression for the exact (nonperturbative) effective action of
$N$=1 supersymmetric gauge theories is proposed, supposing, that all
particles except for the gauge bosons are massive. Analysis of its form
shows, that instanton effects in the supersymmetric theories can lead to
the quark confinement. The typical scale of confinement in MSSM QCD,
calculated from the first principles, is in agreement with the experimental
data. The proposed explanation is quite different from the dual Higgs
mechanism.
\end{abstract}


\section{Introduction}
\hspace{\parindent}

One of the most important unsolved problems of modern theoretical
physics is the explanation of the quark confinement \cite{simonov}.
Usually the confinement is believed to be an essentially nonperturbative
effect. Recently there is an considerable progress in understanding of the
nonperturbative dynamics, caused by the paper \cite{seiberg}. In this
paper the sum of all instanton  corrections for the simplest case of $N$=2
supersymmetric Yang-Mills theory was found. Nevertheless, the model,
considered by Seiberg and Witten is not physical. It is much more
interesting to investigate $N$=1 supersymmetric theories, because indirect
experimental data \cite{exper,dienes} show the presence of $N$=1
supersymmetry in the standard model. That is why below we will consider
this case.

From the other side, the presence of supersymmetry is very important and
essential for the mechanism of confinement, proposed in this paper. This
mechanism is quite different from the dual Higgs mechanism \cite{mandelstam},
which is usually used for the qualitative understanding of this phenomenon.
Actually it is based on an assumption about the structure of effective
potential beyond the perturbation theory and essentially uses the existence
of auxiliary field $D$ in supersymmetric theories.

A great difference between QCD and Grand Unification scales, that is
usually referred as the gauge hierarchy problem is an indirect implication,
that the confinement is induced by instanton effects. Remind, that
two quite different scales naturally arise in instanton calculations
\cite{thooft}, their ratio being proportional to $\exp(-8\pi^2/e^2)$.
However, expressions for the nonperturbative effective action of
$N$=1 supersymmetric gauge theories, proposed in the literature
\cite{affleck,seiberg2,tmf} do not lead to the confinement.

Having analyzed instanton effects in this paper we propose a new
(hypothetical) expression for the effective action (if all particles
except for the gauge bosons are massive), which, in turn, leads to
the confinement, in particular in the $SU(3)$-sector of the Minimal
Supersymmetric Standard Model (MSSM QCD). The typical scale of confinement
can be calculated from the first principles.

In the section \ref{approach} we discuss the structure of instanton
contributions to the effective action and try to find their sum
(i.e. the low energy effective action for $N$=1 supersymmetric Yang-Mills
theory with matter, all particles except for the gauge bosons being
massive) from some general arguments. The superpotential appears to
depend on gauge and auxiliary fields in a very particular way. In the
section \ref{mech} we argue, that such dependence produces the quark
confinement in a sense, that asymptotic color states can not exist. The
investigation shows, that there are two "phases" in the model: a confining
phase at distances, larger, than a critical one and a usual phase at small
distances. The interquark potential is also found in this section. The scale
of confinement is calculated and compared with the experiment in the section
\ref{scale}. The results are briefly discussed in the Conclusion.


\section{Low energy effective action for MSSM QCD}
\hspace{\parindent}
\label{approach}

MSSM QCD is a $N$=1 supersymmetric Yang-Mills theory with the gauge group
$SU(3)$ and six matter supermultiplets. All particles, excluding the
gauge bosons, are massive. In the present paper we will consider a
similar theory with the $SU(N_c)$ gauge group and $N_f$ matter
supermultiplets. It is described by the action

\begin{eqnarray}\label{N1_action}
&&S=\frac{1}{16\pi} \mbox{tr\ Im}\left(\tau \int d^4x\,d^2\theta\ W^2\right)
+\nonumber\\
&&\qquad\qquad\qquad
+\frac{1}{4} \int d^4x\, d^4\theta \sum\limits_{A=1}^{N_f}
\left(\phi^{+}_A e^{-2V}\phi^A + \tilde\phi^{+A} e^{2V} \tilde \phi_A
\right) + S_m,
\end{eqnarray}

\noindent
where the matter superfields $\phi$ and $\tilde\phi$ belong to the
fundamental and antifundamental representations of the gauge group
$SU(N_c)$, and $S_m$ is a sum of all mass terms.

Here we use the following notations:

\begin{eqnarray}\label{notat}
&&\phi(y,\theta)=\varphi(y)+\sqrt{2}\bar\theta(1+\gamma_5)\psi(y)
+\frac{1}{2}\bar\theta (1+\gamma_5)\theta f(y),\nonumber\\
&&\tilde\phi(y,\theta)=\tilde\varphi(y)
+\sqrt{2}\bar\theta(1+\gamma_5)\tilde\psi(y)
+\frac{1}{2}\bar\theta (1+\gamma_5)\theta \tilde f(y),\nonumber\\
&&V(x,\theta)=-\frac{i}{2} \bar\theta \gamma^\mu\gamma_5 \theta
A_\mu(x) + i \sqrt{2}(\bar\theta \theta)(\bar\theta\gamma_5\lambda(x))
+ \frac{i}{4} (\bar\theta \theta)^2 D(x),\nonumber\\
&&W(y,\theta)=\frac{1}{2}(1+\gamma_5)\Big(
i\sqrt{2}\lambda(y) + i\theta D(y)
+\frac{1}{2}\Sigma_{\mu\nu}\theta F_{\mu\nu}(y)+\nonumber\\
&&\qquad\qquad\qquad\qquad\qquad\qquad\qquad\qquad
+\frac{1}{\sqrt{2}}\bar\theta(1+\gamma_5)\theta
\gamma^\mu D_\mu \lambda(y)\Big),\nonumber\\
&&\tau = \frac{\theta}{2\pi}+\frac{4\pi i}{e^2},
\qquad\qquad\qquad\quad\
y^\mu = x^\mu + \frac{i}{2} \bar\theta \gamma^\mu\gamma_5\theta,\nonumber\\
&&
\Sigma_{\mu\nu} = \frac{1}{2}(\gamma_\mu\gamma_\nu - \gamma_\nu\gamma_\mu),
\qquad\qquad
D_\mu = \partial_\mu + i[A_\mu,\ \ ]
\end{eqnarray}

\noindent
where $A_\mu$ is a gauge (gluon) field, $\lambda$ is its spinor
superpartner (gluino) and $D$ is an auxiliary field.

Quarks are built from the fields $\psi$ and $\tilde \psi$ as follows:

\begin{equation}
\Psi = \frac{1}{\sqrt{2}}\Bigg((1+\gamma_5)\psi
+ (1-\gamma_5)\tilde\psi\Bigg)
\end{equation}

\noindent
The scalars $\varphi$ and $\tilde\varphi$ are their superpartners (squarks),
$f$ and $\tilde f$ are auxiliary fields.

Below the supersymmetry is considered to be broken. However, we are
not interested in the concrete mechanism. It is only essential, that
there are some "soft" terms, breaking the supersymmetry, for example,
the gluino mass.

Because the confinement is a low energy phenomenon, we will try to
construct the exact effective action below the thresholds for all
massive particles \footnote{As we will see below, this condition is
not satisfied for $u$ and $d$ quarks. Nevertheless, it is not very
essential, because the mass dependence of the effective action is not
changed above the threshold.}. It should satisfy the following requirements:

{\bf 1.} depend on the original (instead of composite) fields of the theory

{\bf 2.} agree with dynamical (perturbative and instanton) calculations
(This requirement is much more restrictive, than the agreement with the
transformation law of the collective coordinate measure, used in \cite{tmf}).

Of course, the exact effective action can be found only after carrying out
the dynamical calculations and summing the series of instanton corrections.
However, in this paper we present some arguments, which allow to suggest
its form.

First, let us find the general structure of the effective action. We
will assume, that it can be presented as a sum of an expression, invariant
under the supersymmetry transformations, and (mass) terms, which can break
the supersymmetry. We will be interested only in holomorphic part of the
sypersymmetric terms. In order to find it note, that there is a relation
between perturbative and instanton contributions
\cite{thooft,shifman_beta,shifman_review}. In particular,
the requirement of "perturbative" renorminvariance of instanton contributions
allows to construct exact $\beta$-functions of the supersymmetric theories.

Let us express this relation mathematically. Chose a scale $M$ and denote
the value of the coupling constant at this scale by $e$.
Then the one-loop result will be proportional to $-1/4e^2$, while
instanton contributions will be proportional to $\exp(-8\pi^2 n/e^2)$,
where $n$ is a module of a topological number. (One-loop contribution
and instanton corrections are renorminvariant separately.)

The holomorphic part of the {\it perturbative} Wilson effective action
can be written as

\begin{equation}
L_a = \frac{1}{16\pi}\mbox{Im}\,\mbox{tr}
\int d^2\theta\,W^2
\Bigg(\frac{4\pi i}{e^2_{eff}}+\frac{\vartheta_{eff}}{2\pi}\Bigg),
\end{equation}

\noindent
where $e_{eff}$ and $\vartheta_{eff}$ are renorminvariant functions of
fields, the perturbative effective coupling constant and vacuum angle
respectively. Denoting

\begin{equation}
z \equiv \exp\Bigg[2\pi i
\Big(\frac{4\pi i}{e^2_{eff}}+\frac{\vartheta_{eff}}{2\pi}\Big)\Bigg],
\end{equation}

\noindent
the exact effective Lagrangian (including instanton contributions) can
be presented in the following form:

\begin{equation}\label{gs}
L_a = \frac{1}{32\pi^2}\mbox{Im}\,\mbox{tr}\int d^2\theta\ W^2 g(z)
= \frac{1}{32\pi^2}\mbox{Im}\,\mbox{tr}\int d^2\theta\ W^2 \Big(-i\ln z
+\sum\limits_{n=1}^\infty c_n z^n\Big).
\end{equation}

(Let us note, that we did not use so far the constant field approximation,
that is usually assumed for the derivation of exact results.)

In order to find the function $g(z)$, we will use the theorem, that
the conditions

\begin{equation}
g(z) = -i \ln z + \sum\limits_{k=0}^\infty c_n z^n,\qquad
\mbox{Im}\,g(z) > 0;\qquad c_n\in \mbox{Im},
\end{equation}

\noindent
uniquely define its form up to a constant.
(The unequality $\mbox{Im}\,g(z) > 0$ is a requirement of the positiveness
of the effective charge and the condition $c_n\in \mbox{Im}$ follows
from the structure of instanton corrections.) The constant is chosen so
that the effective charge can have arbitrary real values (and, therefore,
there is a point $z$, such that $g(z)=0$). As it was shown in \cite{unique},
these conditions uniquely lead to the following $z$-dependence

\begin{equation}
g(z) = 2\pi\, \tau(z^{-1/4}),
\end{equation}

\noindent
where the function $\tau(a)$ is Seiberg-Witten solution

\begin{equation}\label{tau}
\tau(a) = \left. \frac{da_D(u)}{da}\right|_{\displaystyle u=u(a)},
\end{equation}

\noindent
the functions $a$ and $a_D$ being

\begin{equation}\label{sw}
a(u)=\frac{\sqrt{2}}{\pi}\int\limits_{-1}^{1} dx
\frac{\sqrt{x-u}}{\sqrt{x^2-1}};
\qquad
a_D(u)=\frac{\sqrt{2}}{\pi}\int\limits_{1}^{u} dx
\frac{\sqrt{x-u}}{\sqrt{x^2-1}}.
\end{equation}

Therefore, finally we obtain

\begin{equation}\label{result2}
L_a = \frac{1}{16\pi} \mbox{Im} \int d^2\theta\, W^2 \tau(z^{-1/4}),
\end{equation}

\noindent
where the parameter $z$ can be approximately found in the one-loop
approximation and exactly by the investigation of instanton measure
similar to \cite{shifman_beta,shifman_review}.

Because in the one-loop approximation the $\beta$-function is written as

\begin{equation}
\beta(e)= - \frac{e^3}{16\pi^2} (3N_c-N_f),
\end{equation}

\noindent
the parameter $z$ should be proportional to $M^{3N_c-N_f}$, where $M$ is
an UV-cutoff. The same result can be obtained from the expression for
the instanton measure \cite{cordes} (in a complete agreement with the
above arguments).

The purpose is to construct the effective action in the low-energy limit,
i.e. below the thresholds for all massive particles. The contributions of
massive particles into the running coupling constant are fixed at the
masses and, therefore, their contributions to the parameter $z$ will be
proportional to $M/m$ in a power, defined by the corresponding coefficient
of the $\beta$-function. It is much more difficult to investigate the
contributions of massless gauge fields. Of course, in this case the
coupling constant is not fixed at a definite value and we need to perform
a detailed analysis of the IR behavior of the theory. Note, that we are
interested not in the renormgroup functions but in the effective action,
that can be calculated, for example, in the constant field limit.
Therefore, the contribution of the massless gauge field to the parameter
$z$ will be a function of these fields, instead of masses.

Because $z$ is a scalar, in the constant field limit it is necessary to
find the chiral scalar superfield, which does not contain the derivatives
of $F_{\mu\nu}$ and anticommuting fields in the lowest component (otherwise
all sufficiently large powers of $z$ will be equal to 0 or infinity). The
only superfield $B$ satisfying these requirements is defined as

\begin{equation}\label{r}
B = -\frac{1}{8} \bar D (1-\gamma_5) D (W_a^{*})^2
= (D^a)^2 -\frac{1}{2}(F_{\mu\nu}^a)^2
-\frac{i}{2} F_{\mu\nu}^a \tilde F_{\mu\nu}^a + O(\theta),
\end{equation}

\noindent
where the index $a$ runs over the generators of a gauge group. (At the
perturbative level similar expression was proposed in \cite{shifman_d}).

Taking into account dimensional arguments, we find, that the contributions
of massless gauge fields to the parameter $z$ should be proportional to
$M/B^{1/4}$ in a degree, defined by the corresponding coefficient of the
$\beta$-function.

The one-loop $\beta$-function can be presented as

\begin{equation}
\beta(e) = -\frac{e^3}{16\pi^2} (c_{gauge}+c_\lambda +c_q+c_{sq}),
\end{equation}

\noindent
where we denote

\begin{equation}
c_{gauge}=\frac{11}{3} N_c,\qquad c_\lambda = -\frac{2}{3}N_c,\qquad
c_q = -\frac{2}{3}N_f,\qquad c_{sq}= -\frac{1}{3} N_f
\end{equation}

\noindent
the contribution of Yang-Mills fields (with ghosts), their spinor
superpartners, quarks and squarks respectively. Then, taking into account
the above arguments, we obtain

\begin{equation}\label{z1}
z = e^{-8\pi^2/e^2} M^{3N_c-N_f}
\Bigg(\frac{m_\lambda^{2/3}}{B^{11/12}}\Bigg)^{N_c}
\Big(m_q^{2/3} m_{sq}^{1/3}\Big)^{N_f},
\end{equation}

\noindent
where $m_\lambda$ is a gluino mass, $(m_q)_i^j$ and $(m_{sq})_i^j$ are
mass matrixes of quarks and squarks.

It should be noted, that the equation (\ref{z1}) was found in the one-loop
approximation. Multiloop effects can be included into the {\it Wilson}
effective action \cite{puzzle,shifman_beta,shifman_review}, if we take
into account, that the instanton measure \cite{cordes} contains the factor
$(1/e^2)^{N_c}$, which, certainly, will be present in $z$. Therefore, the
parameter $z$ (up to a constant $C$) is finally written as

\begin{equation}\label{correct_z}
z = C \Bigg(\frac{1}{e^2}\Bigg)^{N_c}
e^{-8\pi^2/e^2} M^{3N_c-N_f}
\Bigg(\frac{m_\lambda^{2/3}}{B^{11/12}}\Bigg)^{N_c}
\Big(\mbox{det}(m_q)_i^j\Big)^{2/3}
\Big(\mbox{det}(m_{sq})_i^j\Big)^{1/3},
\end{equation}

\noindent
where we took into account $SU(N_f)$ symmetry of rotation in the flavor
space and restored indexes of generations.

Let us denote

\begin{equation}\label{LQCD}
\Lambda_c =
\Bigg(\frac{1}{e^2}\Bigg)^{3/11} m_\lambda^{2/11}
\Bigg(
C e^{-8\pi^2/e^2} M^{3N_c-N_f}
\Big(\mbox{det}(m_q)_i^j\Big)^{2/3}
\Big(\mbox{det}(m_{sq})_i^j\Big)^{1/3}
\Bigg)^{3/(11 N_c)},
\end{equation}

\noindent
Then the expression for $z$ can be written in the following form:

\begin{equation}\label{z}
z = \Bigg(\frac{\Lambda_c}{B^{1/4}}\Bigg)^{11 N_c/3}.
\end{equation}

\noindent
As we will see below, $\Lambda_c$ is a typical scale of confinement.


\section{Confinement mechanism}\label{mech}
\hspace{\parindent}

Now, let us note, that the field $D$, although being auxiliary, is
a real quantum field and should be integrated out. In the considered
low energy limit it is necessary to substitute all squarks wave functions
by their vacuum expectation values, which are equal to 0 because the gauge
group $SU(3)$ is not broken. Therefore, the equation of motion for the
field $D$ should be obtained from the action

\begin{equation}
S_W = \frac{1}{16\pi} \mbox{Im}\ \mbox{tr} \int d^4x\, d^2\theta\, W^2
\tau(z^{-1/4}).
\end{equation}

Then, let us note that due to the supersymmetry and (\ref{r}) the field
$D$ is present in $S_W$ only in the following combination

\begin{equation}\label{eq20}
b = (D^a)^2 - \frac{1}{2} (F_{\mu\nu}^a)^2 - \frac{i}{2} F_{\mu\nu}^a
\tilde F_{\mu\nu}^a.
\end{equation}

\noindent
Therefore, as can be easily seen performing the integration over
the  anticommuting variable, the bosonic part of the Lagrangian can
be presented as

\begin{equation}
L_{bose} = \frac{1}{8\pi} \mbox{Im} \Big(b^{*} f(b)\Big),
\end{equation}

\noindent
where $f(b)\equiv \tau(z^{-1/4}(b))$.

In the vacuum state the field $D$ has such value, that $L_{bose}$ is minimal.
If the magnetic field is absent, then $\tilde F_{\mu\nu} F_{\mu\nu}=0$ and
the parameter $b$ can be considered to be real.

\begin{figure}[ht]
\begin{picture}(0,0)(0,0)
\put(2,18){$\mbox{Im}\Big(b f(b)\Big) $}
\put(9,7){1}
\put(7.6,9.2){2}
\put(13.7,3.8){b}
\end{picture}
\epsfxsize15.0truecm\epsfbox{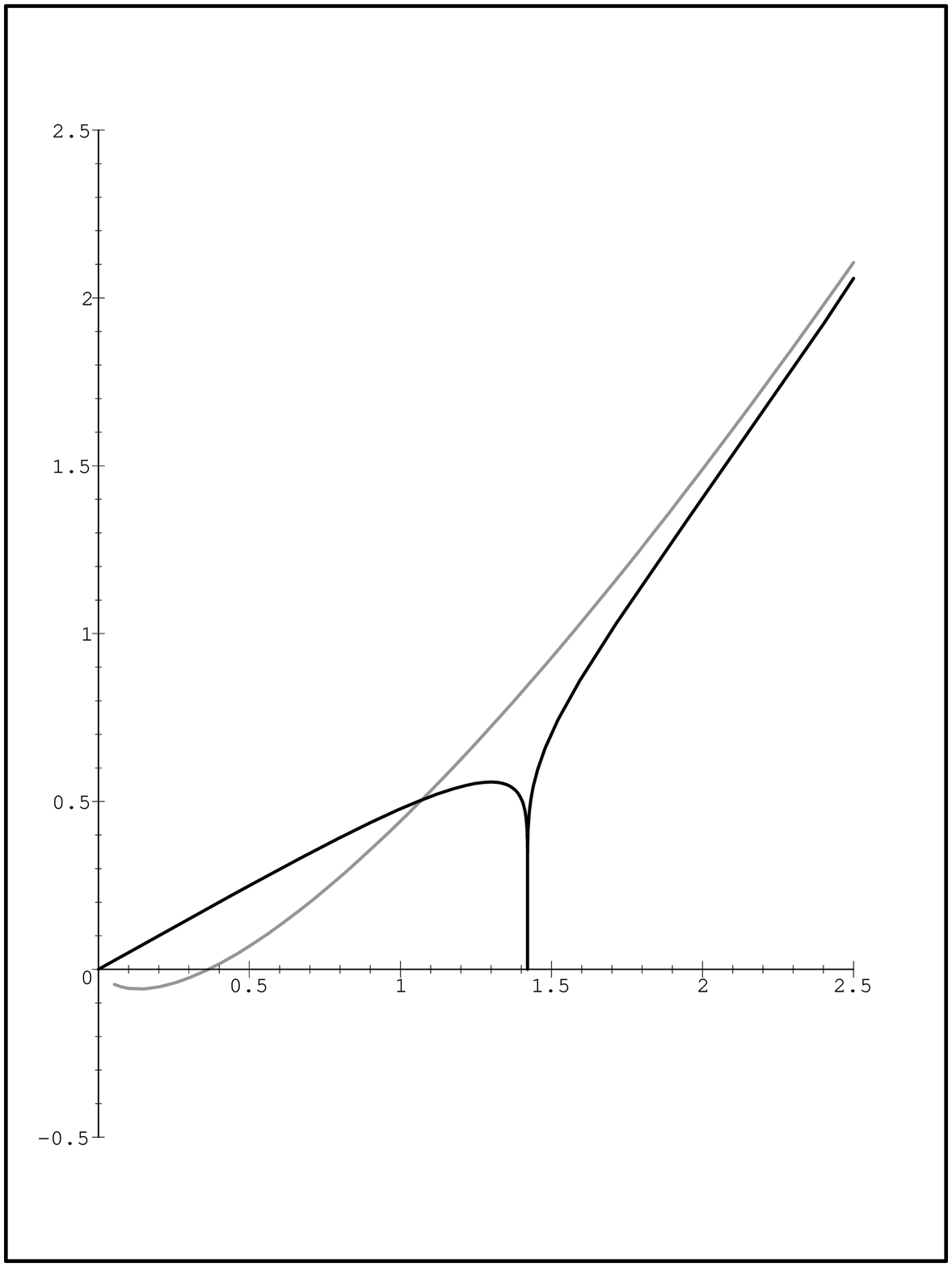}
\caption{The plot of function $\mbox{Im}\Big(b f(b)\Big)$ (curve 1).
For the comparisons we present the plot (curve 2) of the corresponding
perturbative expression.} \label{tau_plot}
\end{figure}

The plot of the function $\mbox{Im}\Big(b f(b)\Big)$ is presented at
Fig.\ref{tau_plot}. (For simplicity we set $\Lambda_{c}=1$ at all figures.)

Using expressions for $a(u)$ and $a_D(u)$ in terms of elliptic functions
\cite{alvares}, it is easy to see, that $\mbox{Im}\Big(b f(b)\Big)$ has
the only extremum $b_0$, satisfying the condition

\begin{equation}\label{extremum}
a(b_0)=z^{-1/4}(b_0)=4/\pi
\end{equation}

In the minimum point $\tau(z^{-1/4}(b_0))=0$, so that there is no
kinetic term for the gauge field in the vacuum state. Therefore,
$A_\mu^a$ is a Lagrange multiplier, which corresponds to the constraint

\begin{equation}\label{curr}
J^a_\mu = 0.
\end{equation}

\noindent
Condition (\ref{curr}) means, that there are no asymptotic color state
and, therefore, leads to confinement of color charges.

At the first sight from the above consideration one can mistakenly
conclude, that color charges can not exist in different space points.
Really, the kinetic term for the Yang-Mills theory is usually written as

\begin{equation}
- \frac{1}{4 e^2} \mbox{tr} F_{\mu\nu}^2,
\end{equation}

\noindent
where $F_{\mu\nu}=\partial_\mu A_\nu - \partial_\nu A_\mu + i[A_\mu,A_\nu]$.
If the corresponding term is absent, the effective charge becomes infinite,
that, in turn, leads to the infinitely strong interaction.

However, investigating the quarks interaction it is necessary to take
into account the cromomagnetic field, produced by quarks, in the simplest
case

\begin{equation}
A_\mu^a = \frac{q^a}{r}\delta_{\mu 0}
\end{equation}

Because in the absence of magnetic field (\ref{eq20}) is real and
can be written as

\begin{equation}
b = (D^a)^2 + (\mbox{\bf E}^a)^2,
\end{equation}

\noindent
the second term is larger than $b_0$ for sufficiently small $r$.
Therefore, it is impossible to reach the minimal value of the function
$\mbox{Im}\Big(b f(b)\Big)$ by any choice of the real field $D^a$. As it
is seen from the presented plot, the function $\mbox{Im}\Big(b f(b)\Big)$
is monotonically growing, so that its minimum corresponds to the minimal
value of $b$, which, certainly, corresponds to $D=0$. Because in this case
evidently $b\ne b_0$, $\tau(b)\ne 0$ and, therefore, gauge field has a
kinetic term and the theory becomes the usual quantum cromodynamics.

Thus, there are two quite different phases in the considered model.
In the first one, corresponding to $b_0>(\mbox{\bf E}_a)^2$, the value
of the auxiliary field $D$ is not equal to 0 (to be exact
$(D_a)^2+(\mbox{\bf E}_a)^2=b_0$), there is no kinetic term for the
gauge field and color charges are confined. In the second phase
$b_0<(\mbox{\bf E}_a)^2$, $D=0$ and the theory can be described by
usual methods. The point of "phase transition" $r_c$ corresponds to

\begin{equation}
b_0 = (\mbox{\bf E}^a)^2 \approx \frac{1}{r^4}
\end{equation}

\noindent
From (\ref{z}) and (\ref{extremum}) we conclude, that

\begin{equation}
(b_0)^{1/4} = \Lambda_{c} \Bigg(\frac{4}{\pi}\Bigg)^{12/(11 N_c)},
\end{equation}

\noindent
so that $\Lambda_{c}$ is really a typical scale of confinement.

Nevertheless, the above picture does not predict the linear growing of the
interquark potential. Instead of it we obtain, that for $r$, larger than
a critical value $r_c$, the potential is infinite. At the small $r$, using
the motion equations, we obtain that

\begin{equation}
\mbox{\boldmath$\nabla$} \Big(\mbox{Im} f_1(1/r^4)
\mbox{\boldmath$\nabla$} U\Big)
= \mbox{const}\ \delta(\mbox{\boldmath$r$})
\end{equation}

\noindent
where $f_1(b) = f(b) + b f'(b)$. Therefore, the potential is given by

\begin{equation}\label{eq30}
U(r)\sim \int\frac{dr}{r^2 \Big(\mbox{Im}f_1(1/r^4)\Big)}
\end{equation}

\noindent
that is actually a Colomn potential, modified by the quantum corrections.

\begin{figure}[ht]
\begin{picture}(0,0)(0,0)
\put(13.6,12.4){$r$}
\put(12.5,12.4){$r_c$}
\put(2,17.7){$U(r)$}
\end{picture}
\epsfxsize15.0truecm\epsfbox{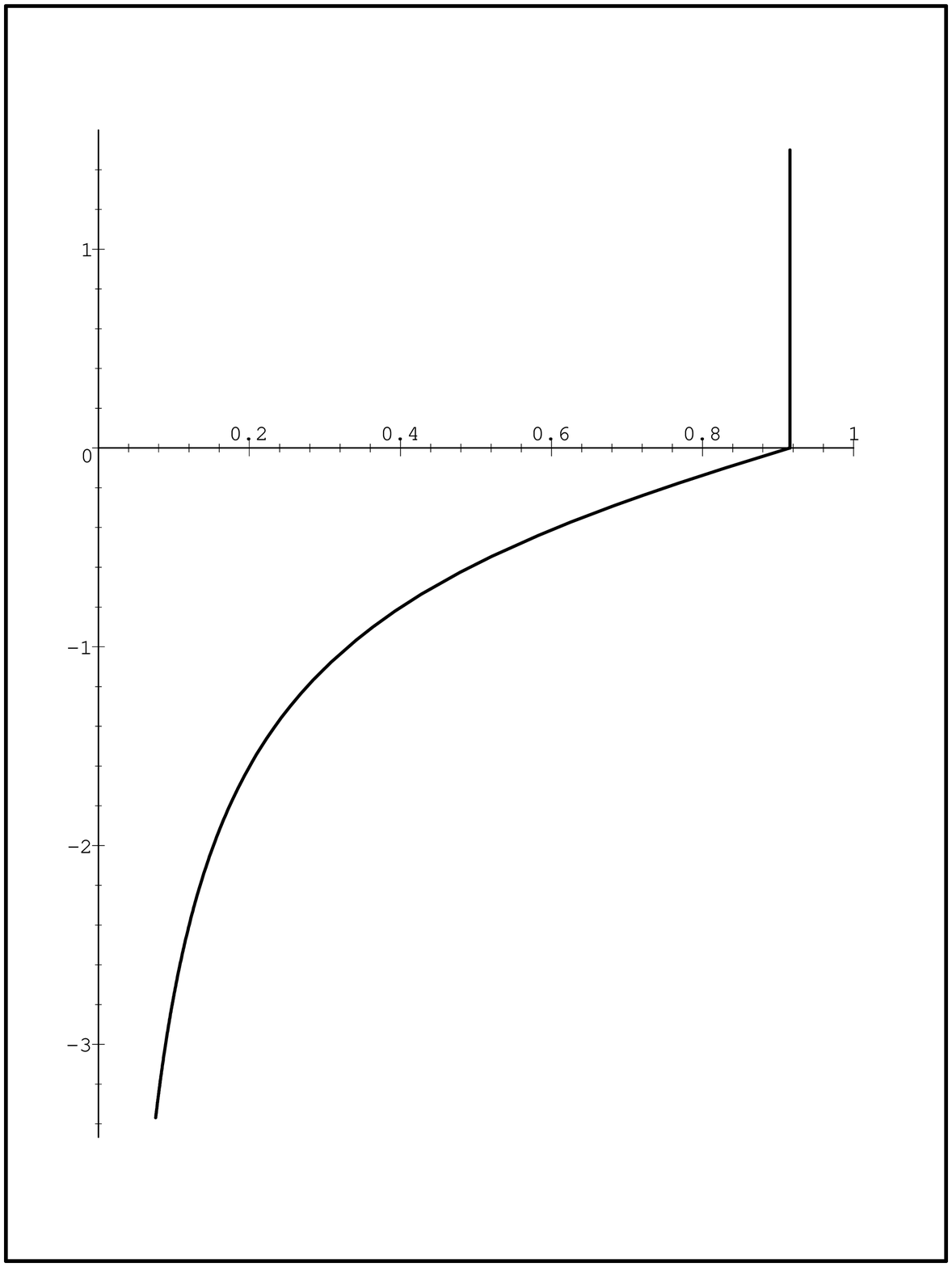}
\caption{The plot of interquark potential $U(r)$ in MSSM QCD.}
\label{potential}
\end{figure}

The plot of potential (\ref{eq30}) is presented at Fig. \ref{potential}.
(As earlier, $\Lambda_{c}=1$ and, moreover, we omit quark charges for
simplicity, that gives $r_c=(b_0)^{-1/4}$.) A normalization constant
is chosen so that $U(r_c)=0$.

\section{Confinement scale}\label{scale}
\hspace{\parindent}

In this section we will investigate only MSSM QCD, which is a particular
case of the considered model corresponding to $N_c=3$ and $N_f=6$.

In principle, due to the renorminvariance, $\Lambda_c$ can be calculated
at any scale, for example, at the Grand Unification scale. In this case
the value of the coupling constant in MSSM should be set to be
$e^2\approx 1/2$ and for the confinement scale from (\ref{LQCD}) we obtain

\begin{equation}
\Lambda_{c} = M e^{-16\pi^2/11}
\left(\frac{m_q^4 m_\lambda^2 m_{sq}^2}{M^8}\right)^{1/11}.
\end{equation}

Unfortunately, we do not know masses of superpartners. Moreover, in this
expression we omit a constant factor, which can be rather large and
slightly change the result. (From the other side, the corresponding
contribution is close to 1, because this factor is in 1/11 degree).

Nevertheless, a rough estimate of $\Lambda_{c}$ can be made. Setting all
masses equal to 10 -- 100 GeV ($M=2\times 10^{16}$ GeV), we obtain that

\begin{equation}
\Lambda_{c}\approx 0.09 - 0.46\ \mbox{GeV}
\end{equation}

\noindent
that is in a good agreement with the experimental data.


\section{Conclusion}
\hspace{\parindent}

In the present paper we propose a mechanism of confinement, which is
quite different from the dual Higgs mechanism \cite{mandelstam}, that
is usually used for understanding of this phenomenon. However, this
mechanism is a rather natural consequence of the structure of exact
effective action (\ref{result2}), with the parameter $z$, given by
equation (\ref{correct_z}).

However, it should be noted, that (\ref{result2}) is only a hypotheses,
that can be confirmed (rejected or modified) only by the perturbative
and instanton calculations, similar to \cite{yung}. In the paper we
presented some facts in its favor, for example, the agreement with
the exact Novikov, Shifman, Vainshtein and Zahkarov $\beta$-function
\cite{shifman_beta} (that can be easily verified) or the positiveness
of the effective charge. Nevertheless, they can not be considered as a
strict proof.

In the mechanism of confinement, proposed in this paper, the two main
point are the most essential: First, it is the existence of auxiliary
field $D$ in supersymmetric gauge theories. Second, we use highly
nontrivial form of the bosonic part of the effective action, which is
obtained by summing a series of instanton corrections.

However, the interquark potential is appeared not to have linear growing
at the large distances. Instead of it there are two different regions
(or "phases"). In the one phase (at distances, larger than a critical one)
the kinetic term for the gauge field is absent and the potential is
infinitely large. In the other phase the theory behaves in a standard way,
the potential being approximately equal to the Colomn one. (Similar
potentials are used in the bag models \cite{close}).

Nevertheless, the proposed mechanism of confinement is not reduced to
the existence of this potential. Actually we predict the absence of color
states at the large distances, that is actually observed in the nature.
Moreover, we automatically obtain the confinement of electric (instead of
magnetic) charges and need not to use duality.

The confinement scale can be calculated from the first principles and
is in a good agreement with the experiment, especially taking into account
the absence of experimental data on the superpartners masses.

\vspace{7mm}

The author is very grateful to P.I.Pronin, O.Pavlowsky, K.Kazakov and
his colleagues from the Institute of Theoretical and Experimental Physics
for valuable discussions and comments. I especially likes to thank
V.V.Asadov for the financial support.


\end{document}